\documentclass[reprint,superscriptaddress,nofootinbib,amsmath,amssymb,aps,pra]{revtex4-1}
\usepackage{graphicx}% Include figure files
\usepackage{dcolumn}% Align table columns on decimal point
\usepackage[inline]{enumitem}
\usepackage{hyperref}% add hypertext capabilities

\usepackage{bm}% bold math
\usepackage{amsmath}
\usepackage[font=small]{caption}
\usepackage[labelformat=empty, position=top]{subcaption}
\usepackage[export]{adjustbox}
\usepackage{blindtext}
\usepackage{amssymb}
\usepackage{newunicodechar}
\usepackage{nameref}

\begin{document}

\title{Characterising Alzheimer’s Disease with EEG-based Energy Landscape Analysis}

\author{Dominik Klepl}
  \affiliation{Centre for Computational Science and Mathematical Modelling, Coventry University, Coventry CV1 2JH, UK 
  }

\author{Fei He}
  \email{Correspondence to: fei.he@coventry.ac.uk}
  \affiliation{Centre for Computational Science and Mathematical Modelling, Coventry University, Coventry CV1 2JH, UK
}

\author{Min Wu}
  \affiliation{Institute for Infocomm Research, Agency for Science, Technology and Research (A*STAR), 138632, Singapore
}

\author{Matteo De Marco}
  \affiliation{Department of Neuroscience, University of Sheffield, Sheffield, S10 2HQ, UK}

\author{Daniel J. Blackburn}
  \affiliation{Department of Neuroscience, University of Sheffield, Sheffield, S10 2HQ, UK}
  
\author{Ptolemaios G. Sarrigiannis}
  \affiliation{Department of Neurophysiology, Royal Devon and Exeter NHS Foundation Trust, Exeter, EX2 5DW, UK}

%\date{}

\begin{abstract}
Alzheimer’s disease (AD) is one of the most common neurodegenerative diseases, with around 50 million patients worldwide. Accessible and non-invasive methods of diagnosing and characterising AD are therefore urgently required. Electroencephalography (EEG) fulfils these criteria and is often used when studying AD. Several features derived from EEG were shown to predict AD with high accuracy, e.g. signal complexity and synchronisation. However, the dynamics of how the brain transitions between stable states have not been properly studied in the case of AD and EEG data. Energy landscape analysis is a method that can be used to quantify these dynamics. This work presents the first application of this method to both AD and EEG. Energy landscape assigns energy value to each possible state, i.e. pattern of activations across brain regions. The energy is inversely proportional to the probability of occurrence. By studying the features of energy landscapes of 20 AD patients and 20 healthy age-matched counterparts, significant differences were found. The dynamics of AD patients’ EEG were shown to be more constrained - with more local minima, less variation in basin size, and smaller basins. We show that energy landscapes can predict AD with high accuracy, performing significantly better than baseline models.
\end{abstract}

\maketitle

\section{Introduction}
\label{sec:introduction}
Alzheimer’s Disease (AD) is a neurodegenerative disorder causing neuronal cell death that leads to dementia, and AD accounts for 70\% of dementia cases. With nearly 50 million patients, it is the most common neurodegenerative disorder in the world \cite{picanco2018review}. Although there is no cure for AD, an early and precise diagnosis can be crucial to prevent or delay the progression of dementia and thus to improve the quality of life of AD patients \cite{cassani2018systematic} . As several new treatments for AD are undergoing evaluation in clinical trials, sensitive, non-invasive and reproducible biomarkers of brain function are urgently required i) to identify and recruit patients in the prodromal phase of the disease, ii) to be implemented as objective outcome measures and iii) to monitor disease progression and potential response to novel treatments \cite{laske2015}. Since current diagnosis mostly relies on neuroimaging scans and invasive tests, both time-consuming and expensive to perform,  low-cost but precise diagnostic methods are required  \cite{cassani2018systematic}. An EEG based biomarker, that can be used as a surrogate endpoint to study the effects of new therapeutic approaches in AD, can become a game-changer in running large pharmaceutical trials. 

Resting-state Electroencephalography (EEG) is a widely and commonly used method in everyday clinical practice, mainly to provide evidence for the diagnosis, classification and management of patients with epilepsy and various other brain disorders (e.g. dementia). EEG is painless, economical, non-invasive, easy to administer and widely available in most hospitals \cite{blackburn2018synchronisation, cassani2018systematic}. EEG measures the changes in electrical currents generated by large populations of cortical neurons. Compared to other neuroimaging methods, EEG provides high temporal resolution on the scale of milliseconds. Recording EEG at rest is advantageous when examining AD patients as it requires little cooperation and is not stressful.

Several characteristic EEG features of AD patients have been documented such as slowing of signals \cite{brenner1988slowing, jeong2004slowing, ghorbanian2015slowing}, reduced complexity  \cite{ahmadlou2010complexity, coronel2017complexity, garn2015complexity} and decreased synchronisation \cite{blackburn2018synchronisation, dauwels2010synchronisation, vysata2015synchronisation}. However, previous work mainly analysed individual channels, or pairs of channels  \cite{brenner1988slowing, jeong2004slowing, ghorbanian2015slowing, blackburn2018synchronisation, dauwels2010synchronisation, vysata2015synchronisation}. In comparison, we use pairwise analysis only as a basis for estimating global properties of the system (or the region of interest), e.g. group of 10 channels. Rather than claim that our approach achieves better classification accuracy, the focus of this study is to characterise AD from a novel perspective – the global system, network and information-theoretic energy viewpoint.

Although neuron loss and an accumulation of protein aggregates, the beta-amyloid plaques, are a diagnostic hallmark of Alzheimer's disease, 
the exact cause of neurodegeneration remains to be elucidated. The early 
loss of neocortical synapses appears also to play a key role in brain network dysfunction and correlates well with various cognitive deficits   \cite{dennis2014connectivity,terry2000celldeath} and  altered functional connectivity \cite{dennis2014connectivity, fu2019connectivity, vysata2015synchronisation}. Thus, AD can be viewed as a network disorder.

Aiming to characterise brain network organisation in AD, this study models the EEG using the energy landscape (EL). This approach allows quantifying global characteristics of a dynamic complex system such as the brain. As a powerful emerging method in neuroscience, EL conceptualises the brain signals as a network of distinct states. Each state has an energy which refers to the negative log probability that the system is in a given state \cite{kang2019graph}. Note that energy is an information-theoretic measure and has no link to the physical concept of energy. Pairwise maximum entropy model (pMEM) is used to estimate the energy of each state. Schneidman et al. \cite{schneidman2006cells} demonstrated the first application of pMEM and EL to study the dynamics of neural networks on a single-cell level. More recently, EL has been applied to various neuroimaging data such as fMRI \cite{ezaki2018landscape, kang2019graph, ashourvan2017modules, gu2018energy} and MEG \cite{krzeminski2020meg}. To our knowledge, this study is the first attempt to extend the method to EEG analysis.

The EL method was mostly used to analyse fMRI data, which characteristics differ from those of EEG. The spatial resolution of fMRI is considerably higher than EEG \cite{bunge2009techcomparison}, which allowed previous fMRI work to focus on the analysis of specific brain networks with rather accurate anatomical and functional definitions. Therefore, in previous studies, the brain regions were manually selected \cite{ezaki2018landscape, kang2019graph, ashourvan2017modules, gu2018energy, krzeminski2020meg, schneidman2006cells}. However, EEG does not allow accurate monitoring of specific networks as it measures the activity of macro-regions. Unless a structural scan of the brain is available along with EEG data, which would allow source localisation, it is impossible to perform the analysis on the same level as with fMRI. We overcome the low spatial resolution of EEG by adopting a sensor level data-driven approach. We use two channel selection methods to optimise predictive accuracy and channel activity while taking into account the nonlinear relationships between the EEG channels (see Methods for details).

The second difference between fMRI and EEG data is the temporal resolution. While fMRI measures the neural activity on a scale of seconds, EEG can measure on a scale of milliseconds \cite{bunge2009techcomparison}. However, this difference does not seem to pose any methodological issues for adapting EL to EEG data.

It has been shown that EL constructed using pMEM is a powerful, yet relatively simple method to study network connectivity, outperforming all traditional connectivity measures \cite{watanabe2013pmem}. It has been used to study resting state dynamics \cite{ezaki2018landscape}, multi-stability and state transitions during rest \cite{kang2019graph}, module dynamics \cite{ashourvan2017modules} and juvenile myoclonic epilepsy \cite{krzeminski2020meg}. However, there is a lack of studies investigating the EL characteristics of brain disorders, except for epilepsy. 

This study aims to quantify the differences in EL of AD patients compared to age-matched healthy participants and use these differences to automatically classify patients with AD. pMEM is used to estimate the EL, as the concept of energy is defined using this model, i.e. the probability of each state is given by baseline activation of each channel and pairwise linear interactions between channels. First, 10 channels are selected using a combination of filter and wrapper channel selection methods. Then, the EEG signals must be binarised \cite{ezaki2018landscape}. However, as EEG is not stationary, we use a time-varying threshold for the binarisation instead of the traditional fixed threshold. The window size is selected so that the averaged predictive performance of all models is maximised. We proceed with demonstrating that the models trained using EL features perform significantly better than baseline models trained on connectivity parameters of the pMEM. Both of these models are trained using support vector machine (SVM) with radial basis kernel, 10-fold cross-validation and identical hyper-parameter tuning to ensure fairness of comparison. Finally, we show the differences between AD and HC in terms of features of EL.

Unlike other methods, the EL can quantify the nonlinear dynamics of a high-dimensional system such as the brain only in terms of 2 sets of pMEM parameters, the baseline activity and the pairwise interactions of channels. The proposed method can be viewed as an extension of analysis of functional connectivity networks as it models the output of the system, i.e. the emitted electric signals, as a product of a weighted network of channels that each have their own independent activity and linear pairwise interactions. Hence, the energy landscape analysis based on pMEM can provide a more global picture of the properties (i.e. distributions and energy) of all the brain ‘states’, rather than the simple pairwise, local properties from the common functional connectivity analysis. In the current study, we extrapolate the established method of EL from fMRI to EEG.

\begin{figure*}[!ht]
    \centering
    \includegraphics[width=0.85\textwidth]{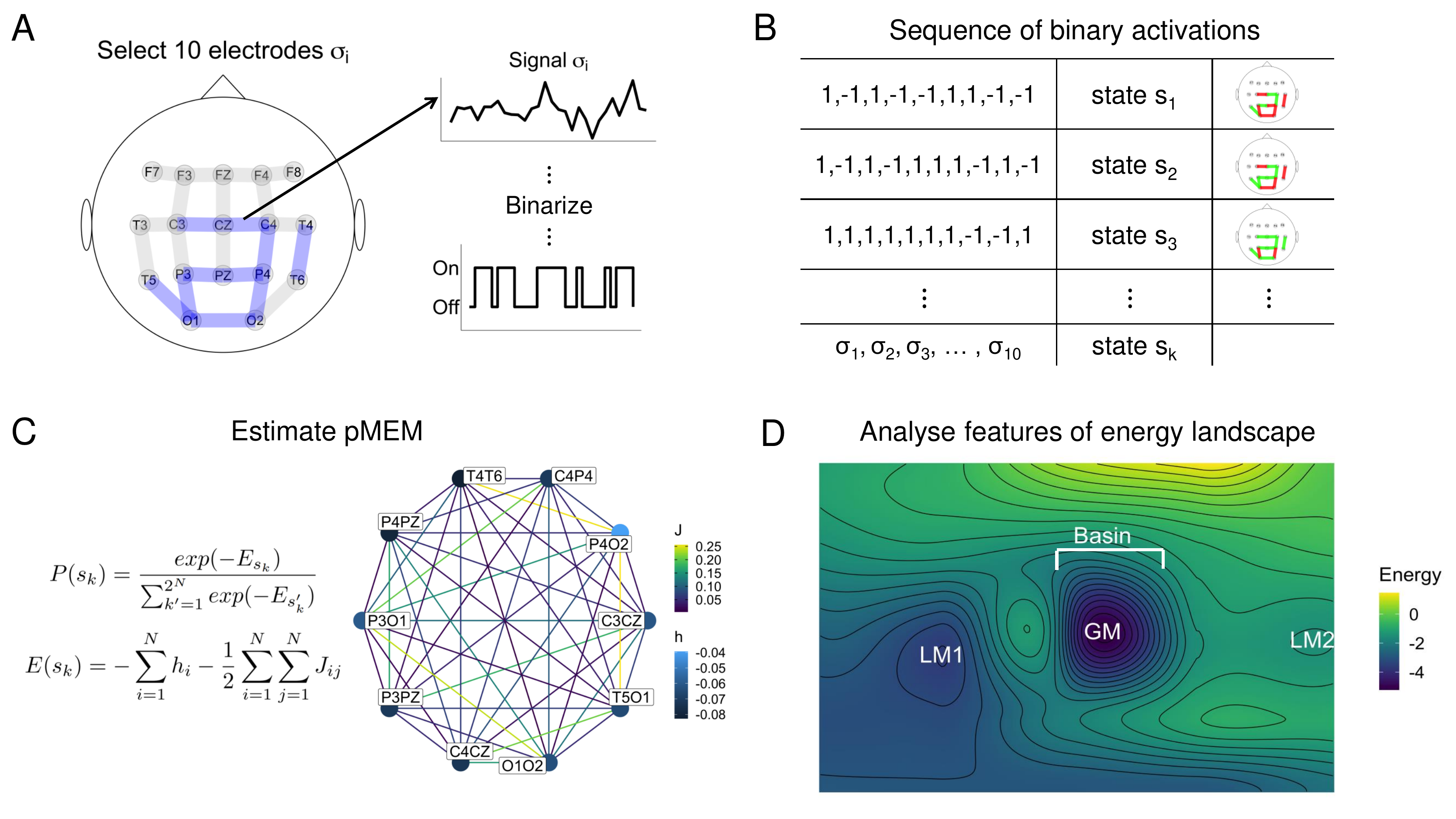}
    \caption{A conceptual schematic of implementing energy landscape to EEG data. A) 10 EEG channels are selected using a combination of filter and wrapper methods. The continuous signal from each channel is then thresholded. Values above mean become 1 and values below mean become -1. B) The binarised dataset. Each row represents a brain state at given timepoint. C) Energy of each brain state is estimated using the pairwise maximum entropy model (pMEM). pMEM estimates the energy using J and h parameters representing the functional connectivity between channels and base rate of activation, respectively. D) Features of a constructed landscape can be extracted. We count the number of local minima (LM), the standard deviation of basin sizes, mean energy difference between LM and global minimum (GM), and simulate duration in the basin of GM.}
    \label{fig:overview}
\end{figure*}

\section{Data}
This study uses EEG recordings collected from 20 AD patients and 20 healthy participants (HC) under 70. In addition, we have a separate dataset of 9 AD and 10 HC, who are all above 70. This dataset is used to validate the results obtained with the under 70 dataset. For a detailed description of the EEG electrode configuration, experimental design and confirmation of diagnosis, see \cite{blackburn2018synchronisation}. All AD participants were recruited in the Sheffield Teaching Hospital memory clinic, which focuses mainly on young-onset memory disorders. The participants were diagnosed with AD between 1 month and 2 years prior to data collection. Their diagnosis was determined using evidence from medical history, the battery of psychological tests and neurological and neuroradiological examinations. All of them were in the mild to moderate stage of the disease at the time of recording with the average Mini Mental State Examination score of $20.1 (sd = 4)$. To eliminate alternative causes of dementia, high resolution structural magnetic resonance imaging (MRI) scans of all patients were acquired. Age and gender matched HC participants (neuropsychological tests and structural MRI scans were normal) were also recruited.

All EEG data were recorded using an XLTEK 128-channel headbox with Ag/AgCL electrodes placed on the scalp and sampling frequency of 2 kHz. A modified 10-10 overlapping a 10-20 international system of electrode placement was adopted. A referential montage with a linked earlobe reference was used. The recordings lasted 30 minutes, during which the participants were instructed to rest and not to think about anything specific. Within the 30 minutes recording, there were two-minute-long epochs during which the participants had their eyes open (EO) or closed (EC).

All the recordings were reviewed by an experienced neurophysiologist on the XLTEK review station with time-locked video recordings (Optima Medical LTD). For each participant, 3 12 second long artefact-free EO and EC epochs were isolated. Finally, to avoid volume conduction effects related to the common reference electrodes, the following 23 bipolar channels were created: F8–F4, F7–F3, F4–C4, F3– C3, F4–FZ, FZ–CZ, F3–FZ, T4–C4, T3–C3, C4–CZ, C3–CZ, CZ–PZ, C4–P4, C3–P3, T4–T6, T3–T5, P4–PZ, P3–PZ, T6– O2, T5–O1, P4–O2, P3–O1 and O2–O1. This study was approved by the Yorkshire and The Humber (Leeds West) Research Ethics Committee (reference number 14/YH/1070). All participants gave their informed written consent.

\subsection{Data preprocessing}
Typically, EEG is susceptible to noise, which can originate from external sources such as power line, participant’s movements, and muscle contraction artefacts. In this study, Fourier transform filter (FTF) is used to denoise the signals. The removed frequencies were informed by the findings of meta-analysis on resting-state EEG of AD patients \cite{cassani2018systematic}. Following frequencies were removed: 0-0.5 Hz relating to slow artefacts and eye-blinks, 50 Hz relating to the power line noise and 100 Hz and above, which could result from muscle movement. As a result, the cleaned signals are within 0.5 and 100 Hz.

Then, we separate the EEG signals into frequency bands (further as bands) to analyse the data in finer detail. We create 6 bands using the FTF: delta ($<$ 4 Hz), theta (4 - 7 Hz), alpha (8 - 15 Hz), beta (16-31 Hz), gamma (32 - 100 Hz), full (0.5 - 100 Hz). Thus, the EEG signals are split into 6 time-series.

Next, the Hilbert transform was applied to obtain the analytic signal. The amplitude envelope is computed as the absolute value of the analytic signal. Thus, we study the differences in amplitude correlation of EEG signals. Amplitude is binarised as it has a meaningful interpretation \cite{krzeminski2020meg}, i.e. low amplitude implies low or no activity within a given region, and thus the region being denoted as ”off” and vice versa.

\section{Methods}
\label{sec:methods}

In this study, we analysed the EL of resting EEG signals. The preprocessed data are used to select 10 channels. The channel selection uses entropy-based method and backward elimination jointly.

EL models the probability of occurrence of a state by using the concept of energy. The higher energy of a state, the lower is the estimated probability of the system being in the given state. 
EL is computed via pMEM, estimated using pseudo-likelihood maximisation method \cite{ezaki2018landscape}. The pMEM is fitted to binarised EEG signals. The goodness-of-fit is evaluated using an accuracy index $r$ \cite{ezaki2018landscape, krzeminski2020meg}. The advantage of EL analysis is tested by comparing the predictive power of the energy of all states to baseline models based on connectivity. Multiple features of ELs are extracted in order to analyse the differences between AD and HC.

\subsection{Channel selection and signal binarisation}
The number of channels needs to be reduced since the computational cost of constructing EL increases exponentially with number of channels. The channels are selected in an automatic, data-driven manner. The channel selection optimises predictive accuracy and channel activity while accounting for nonlinear relationships between the channels. Technically, a filter and a wrapper selection methods are implemented \cite{alotaiby2015review}. Channel selection is performed for each frequency band separately.

Predictive accuracy is optimised as we aim to investigate the predictive power of EL. A wrapper method using backward elimination with SVM is implemented to select these channels. SVM with 5-fold cross-validation is used as it can detect both linear and nonlinear patterns in the data. The folds are created to keep data from the same participant within the same fold to prevent information leakage. The importance of channel i is evaluated using permutation importance  \cite{fisher2019permutation}:
\begin{equation}
    F_i = A^o / A^p,
\end{equation}
where $A^o$ is the accuracy of the trained model and $A^p$ is the accuracy obtained with values of channel $i$ randomly shuffled. The backward elimination is initialised with all channels being used to train the SVM. In each iteration, the channel with the lowest importance is removed, and the model is re-trained. This is repeated while the cross-validated accuracy increases. In such a way, we obtain a set of channels $S^{wrap}$ and their respective permutation importance.

Next, channel activity is optimised as constructing EL of channels with high activity is desirable as it ensures that state-transitions are maximised. An entropy filter method was selected for this purpose with entropy of channel $X_i$ given by
\begin{equation}
    H(X_i) = - \sum_{j=1}^{N} P(x_j) log P(x_j),
\end{equation}
where $P$ is the probability distribution of $X_i$ obtained with kernel density estimation. We obtain a set of 15 channels $S^{filter}$ that is ordered by entropy values in descending order.

10 of the channels selected jointly by both methods are chosen as the final channels for each band, i.e. $S^{wrap} \cap S^{filter}$. If there are more than 10 common channels, the channels in $S^{wrap}$ with the highest permutation importance are selected. If there are less than 10 common channels, the remaining channels in $S^{wrap}$ are selected.

Next, the continuous EEG signals from selected channels are binarised so that values above mean become 1, and -1 otherwise \cite{ezaki2018landscape, kang2019graph, krzeminski2020meg}. As a result, for each EEG recording with N electrodes we obtain a sequence of binary signals $\{\sigma_i(1),..., \sigma_i(T)\}$, where $T$ is the number of the recorded samples. $\sigma_i (t) = 1$ means that the brain region measured by the electrode $i$ is active at time $t$. Thus, the state of the system at time $t$ is represented by an $N$-dimensional vector $s(t) = (\sigma_1,...,     \sigma_N)\in\{-1,1\}^N$. There are $2^N$ possible states $s_k (k = 1...,2^N)$. Previously, the sample mean computed from the whole time series had been used \cite{ezaki2018landscape}. However, EEG signals are non-stationary; thus, we use a time-varying threshold to perform the binarisation.

\subsection{Pairwise maximum entropy model}
The pMEM describes the probability of each state, using Boltzmann distribution and two parameters, $h$ and $J$. $h_i$ quantifies the baseline activity of  $ith$ electrode while $J_{ij}$ quantifies the interaction between $ith$ and $jth$ electrodes.

We start by calculating the empirical frequency of each state $s_k$, $P_{emp}(s_k)$, then calculate the empirical activation rate of each electrode $\langle \sigma_i \rangle_{emp}$ and the pairwise co-occurrence of any two electrodes $\langle \sigma_i \sigma_j \rangle_{emp}$ from
\begin{equation}
    \langle \sigma_i \rangle_{emp} = \frac{1}{T} \sum_{t=1}^T \sigma_i(t)
\end{equation}
\begin{equation}
    \langle \sigma_i \sigma_j \rangle_{emp} = \frac{1}{T} \sum_{t=1}^T \sigma_i(t)\sigma_j(t).
\end{equation}

Next, we fit Boltzmann distribution to $P_{emp}(s_k)$
\begin{equation} \label{prob_pMEM}
    P_{pMEM}(s_k|h,J) = \frac{\exp[-E(s_k|h,J)]}{\sum_{k'=1}^{2^N}\exp[-E(s_{k'}|h,J)]},
\end{equation}
where $E(s_k)$ denotes the energy of the state $s_k$, given by
\begin{equation} \label{energy}
    E(s_k) = - \sum_{i=1}^N h_i \sigma_i(s_k) - \frac{1}{2} \sum_{i=1}^N\sum_{\substack{j=1 \\ j\neq i}}^N J_{ij} \sigma_i(s_k) \sigma_j(s_k),
\end{equation}
with $\sigma_i(s_k)$ being the $ith$ element of the state $s_k$. $h$ and $J$ are the parameters of the model, described above, that are to be estimated from the data. Based on the maximum entropy principle, the parameters $h$ and $J$ are selected so that $\langle \sigma_i \rangle_{emp} = \langle \sigma_i \rangle_{mod}$ and $\langle \sigma_i \sigma_j \rangle_{emp} = \langle \sigma_i \sigma_j \rangle_{mod}$. The activation rate $\langle \sigma_i \rangle_{mod}$ and pairwise co-occurrence $\langle \sigma_i \sigma_j \rangle_{mod}$ predicted by the model are given by
\begin{equation}
    \langle \sigma_i \rangle_{mod} = \sum_{k=1}^{2^N} \sigma_i(s_k) P_{pMEM}(s_k|h,J)
\end{equation}
\begin{equation}
    \langle \sigma_i \sigma_j \rangle_{mod} = \sum_{k=1}^{2^N} \sigma_i(s_k) \sigma_j(s_k) P_{pMEM}(s_k|h,J).
\end{equation}

We estimate $h$ and $J$ from the data using the pseudo-likelihood maximisation approach \cite{ezaki2018landscape} with a learning rate $0.1$, a stopping criterion of $5 \times 10^{-6}$ and the maximum number of iterations $1 \times 10^5$.

\subsection{Goodness-of-fit}
Previous studies report several metrics to evaluate how well the pMEM approximates the empirical data and what is the contribution of including pairwise interactions in the model.  We use accuracy index $r$ \cite{ezaki2018landscape, krzeminski2020meg, watanabe2013pmem} given by:
\begin{equation}
    r = (D_1 - D_2)/ D_1,
\end{equation}
where $D_2$ represents the Kullback-Leibler divergence between the probability distribution estimated with pMEM (\ref{prob_pMEM}) and the empirical data, which is given by:
\begin{equation}
    D_2 = \sum_{k=1}^{2^N} P_{emp}(s_k) \log_2 \Big[\frac{P_{emp}(s_k)}{P_{pMEM}(s_k)}\Big].
\end{equation}

$D_1$ represents the Kullback-Leibler divergence between the probability distribution estimated with independent maximum entropy model (iMEM) and empirical data. iMEM does not consider any pairwise interactions, i.e. $J = 0$.

Thus, $r$ represents the contribution of pairwise interactions. $r = 1$ when the pMEM predicts the empirical distribution without error, and $r = 0$ when including the pairwise interactions does not contribute to the prediction of empirical distribution.

\subsection{Constructing energy landscape}
After the pMEM model is fitted, (\ref{energy}) can be used to obtain the energy of each of $2^N$ possible states $s_k$. EL can then be framed as an undirected network of states where edges denote transition between two states. We assume a gradual transitions between the states so an edge connects two states with hamming distance = 1. For example, consider a network where $N=3$, there is connection between states [1,1,1] and [-1,1,1] but states [1,1,1] and [-1,-1,1] are not connected.
Thus we can represent the states of EL as an adjacency matrix:
\begin{equation} \label{adjacency}
    A(i,j) = 
    \begin{cases}
    1 & \text{if $D_H=1$}\\
    0 & \text{otherwise}
    \end{cases},
\end{equation}
where $D_H$ is the Hamming distance between states $i$ and $j$.

Local minimum (LM) of EL is a state whose energy is lower than all of its neighbouring states. Thus, we exhaustively compare the energy of all states and their neighbours. We also identify the global minimum (GM).

The number of LM and mean energy difference between GM and LMs are then further analysed. Each LM can be viewed as an attractor with a field of attraction, i.e. basin. We implement an algorithm for computing the basin size \cite{watanabe2013pmem}. One state is selected, and we move through the EL by moving to a lower-energy neighbour until an LM is reached. The initial state is then assigned to the basin of the LM, where the algorithm stopped. This is repeated for all states. The standard deviation of the basin sizes of all LMs is then computed.

Finally, we use the EL as a generative model to simulate a sequence of state transitions. To simulate the transitions in the EL, we use Markov chain Monte Carlo sampling. The sampling is initialised at random state. In each iteration one of neighbours $s_{c'}$ of the current state $s_{c}$ is selected with probability $1/N$. If $E(s_c)<E(s_{c'})$ then the system moves to the selected state $s_{c'}$ with probability of $\exp[E(s_c)-E(s_{c'})]$. Otherwise, the probability of moving is 1.

We obtain 24 000 samples. The first 2000 samples are removed to minimise the potential effect of the starting state. Using the samples, time the system spends in the GM basin is computed and further analysed.

\subsection{Methods summary}
Besides the parameters of pMEM, we have two additional parameters to select, i.e. sampling frequency and window size. The original sampling frequency of the data is 2000 Hz. However, such a high sampling frequency may be unnecessary as the same amount of information is often retained with smaller sampling frequencies while reducing the computational load. 4 sampling frequencies were tested: 500, 1000, 1500 and the original 2000 Hz.

Several window sizes were also tested for each sampling frequency: 0.1, 0.2, 0.3, 0.4, 0.5, 1, 1.5, 2, 2.5, 3, 3.5 and 4 seconds. As the window sizes are measured in seconds, the actual window size scales by the given sampling frequency.

For all combinations of sampling frequencies, window sizes, bands and conditions, pMEMs are estimated, and two machine learning models are trained: using values of J parameters of pMEM (Connectivity) and the energy values of all states (Energy). A radial-basis kernel SVM and 10-fold cross-validation is used with samples from the same patient being kept within the same fold. PCA is used to reduce the dimensions while preserving 95\% of the variance. PCA is computed for each iteration of cross-validation using only training data.

In order to select the sampling frequency and window size, an average of the area under the receiver operating characteristic curve (AUC) values is computed. The analysis is further performed only on the EL obtained using the selected sampling frequency and window size.

The goodness-of-fit of the pMEM models is then evaluated to verify that any detected differences between groups are not due to differences in the goodness-of-fit of pMEM. ANOVA is used to compare between the groups, bands and conditions.

Next, we test whether energy models perform better than a baseline model using ANOVA. The connectivity model is used as a baseline, and we argue that the EL should not be analysed unless it performs significantly better than the baseline.

The differences in the features extracted from the ELs are analysed. These tests can be viewed as an interpretation of the differences in ELs learned by the models trained on energy values. These features are compared between groups, conditions and bands using an ANOVA. Any significant differences are tested with Tukey’s post hoc tests.

Finally, we measured correlations between energy values of brain states and Mini-Mental State Exam (MMSE) score of AD patients to reveal the potential of EL to deliver a better diagnosis. Kendall’s $\tau$ correlation is used as MMSE is a scale, and the p-value is FDR-adjusted.

All experiments were done using R 4.0.4, and the code is available at \url{https://github.com/dominikklepl/AD-energy-landscape}.

\section{Results}

\subsection{Channel selection}
\begin{figure}[!ht]
    \centering
    \includegraphics[width=\linewidth]{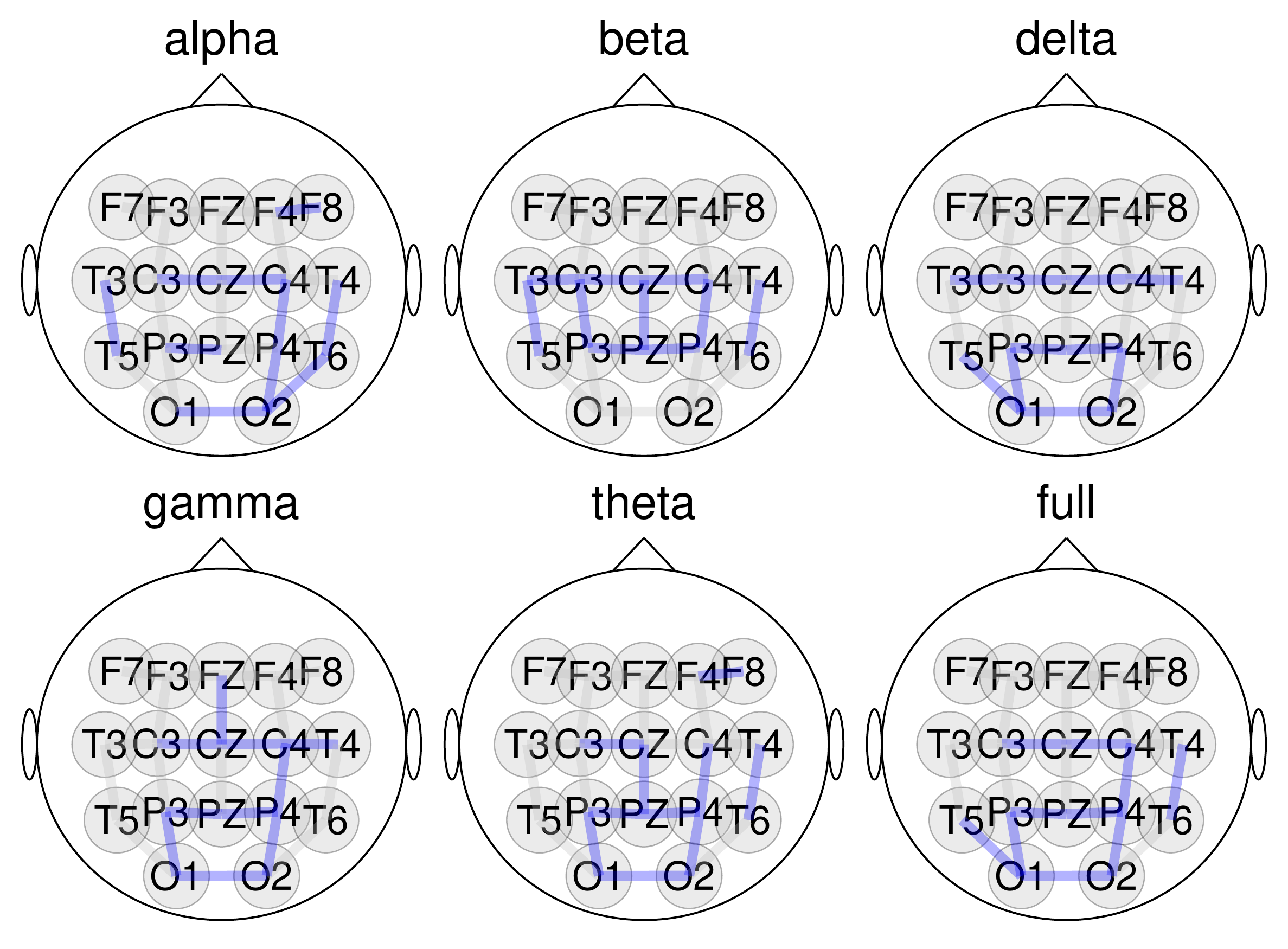}
    \caption{Selected channels by the overlap of the entropy filter and the SVM wrapper methods.}
    \label{fig:selection}
\end{figure}

Fig. \ref{fig:selection} shows the selected channels in each band. The selected channels show many similarities across all bands, with posterior and central channels more likely to be selected. There are only a few deviations from this pattern, such as in alpha and theta bands, where a few frontal channels were also selected.

\subsection{Selection of sampling frequency and window size}

\begin{figure}[!ht]
    \centering
    \includegraphics[width=0.9\linewidth]{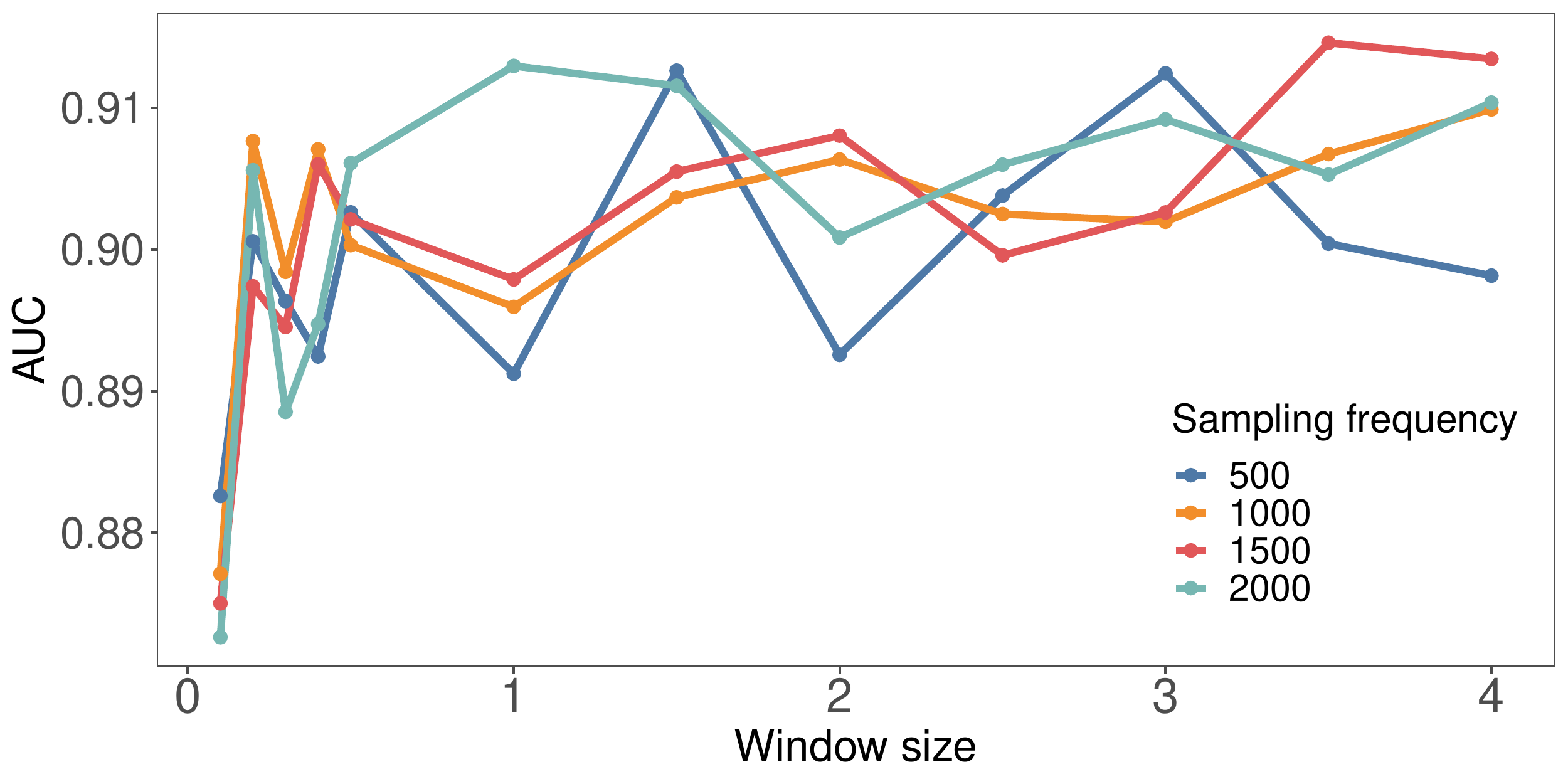}
    \caption{Cross-validated performance of models (in AUC) across different sampling frequencies and window sizes. The performance for each parameter combination is averaged over all bands, conditions and types of machine learning models.}
    \label{fig:selecting-parameters}
\end{figure}

The predictive performance across multiple sampling frequencies and window sizes was tested (Fig. \ref{fig:selecting-parameters}). The AUC was maximised at a sampling frequency of 1500 Hz and a window size of 3.5 seconds (=5250 samples). The remaining results refer to models with these parameters.

\subsection{Goodness-of-fit of pMEM}
pMEM was fitted for each participant separately using the gradient descent based parameter updating described in the previous section. The goodness-of-fit of the pMEM was evaluated using $r$. ANOVA was used to test for differences in $r$. No significant difference between groups was found $(F(1,1440)=0.819, p=0.366)$. Significant main effects of band $(F(5,1440)=468.383, p<0.001)$ and condition $(F(1,1440)=12.413, p<0.001)$ were found, suggesting that the information carried by different bands and conditions influences the goodness-of-fit of pMEM.

The pMEM estimates the probability distribution with relatively high accuracy with mean $r = 0.499 (sd = 0.176)$.

\subsection{Performance against baseline}
\begin{figure}[!ht]
    \centering
    \includegraphics[width=\linewidth]{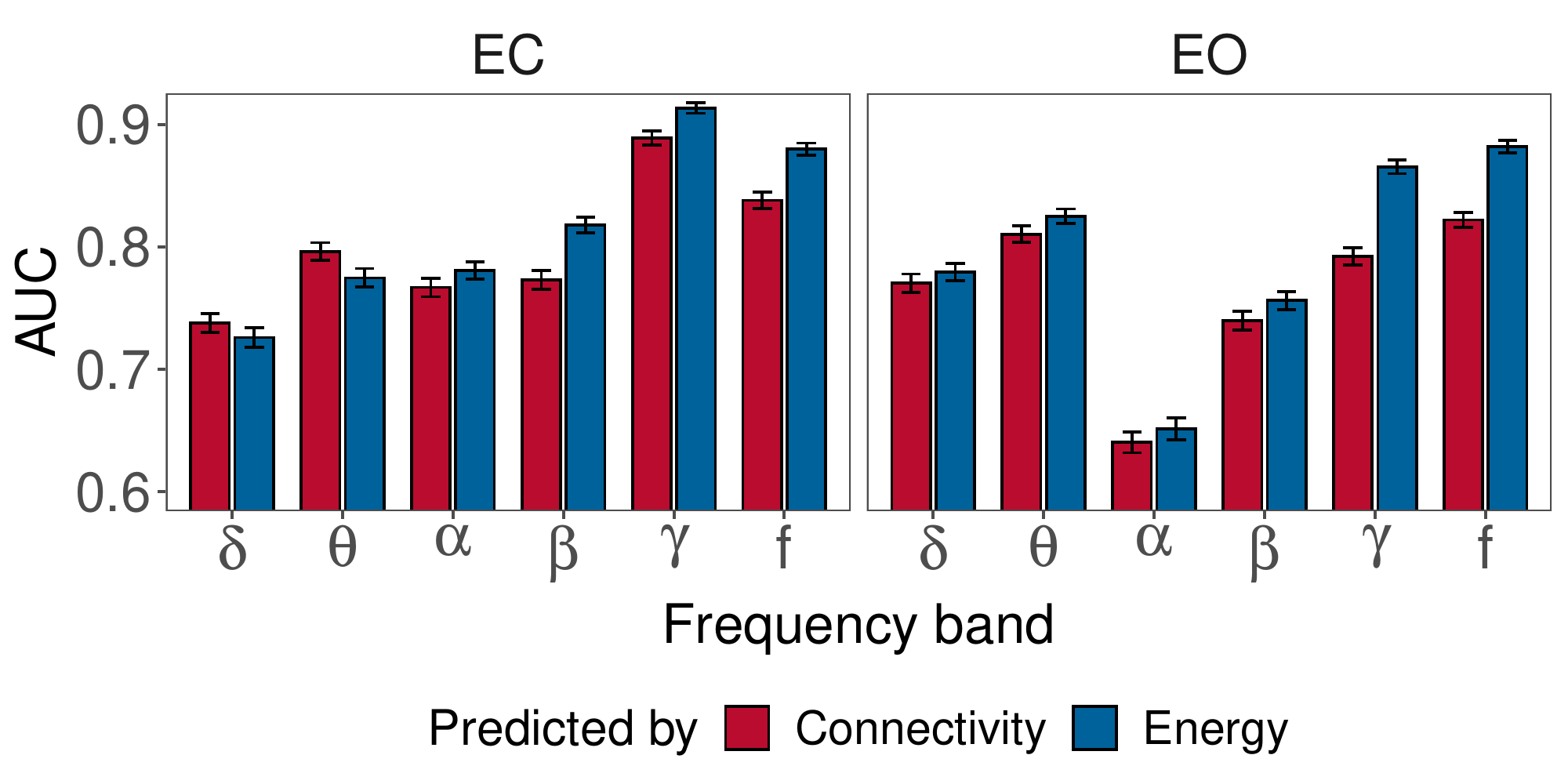}
    \caption{Comparison of performance of energy based models and baseline models. Bars show mean value and errorbars show the 95\% confidence interval of mean.}
    \label{fig:energy-connectivity}
\end{figure}

Samples from the performance distributions of both models were tested for differences between the two types of models. There are significant main effects of model type $(F(1,240)=229.813,p<0.001)$, band $(F(5,240)=1089.788,p<0.001)$, and condition $(F(1,240)=420.935,p<0.001)$. A significant interaction of model type, band and condition was found $(F(5,240)=14.733,p<0.001)$.
Post hoc tests reveal no significant difference in $\delta$ $(F(1,20)=3.04,p=0.162)$ and $\alpha$ $(F(1,20)=3.054,p=0.162)$ bands during EO condition and $\delta$ $(F(1,20)=4.276,p=0.117)$  during EC condition. Connectivity-based models perform significantly better in $\theta$ $(F(1,20)=16.504,p<0.001)$ band during EC condition. In the remaining band-condition combinations, the energy-based models perform significantly better (Fig. \ref{fig:energy-connectivity}). In EC condition that is: $\alpha$ $(F(1,20)=7.069,p=0.032)$, $\beta$ $(	
F(1,20)=74.313,p<0.001)$, $\gamma$ $(F(1,20)=39.355,p<0.001)$ and full frequency $(F(1,20)=93.933,p<0.001)$. In EO condition that is: $\theta$ $(F(1,20)=10.218,p=0.006)$, $\beta$ $(	
F(1,20)=9.088,p=0.015)$, $\gamma$ $(	
F(1,20)=249.048,p<0.001)$ and full frequency $(F(1,20)=210.633,p<0.001)$.

%\begin{figure}[!ht]
%    \centering
%    \includegraphics[width=\linewidth]{best_ML_energy.pdf}
%    \caption{Performance of energy-based models trained using leave-one-patient-out cross-validation.} 
%    \label{fig:energy-performance}
%\end{figure}

\subsection{Differences in energy landscapes}

\begin{figure*}[!ht]
    \centering
    \includegraphics[width=0.8\linewidth]{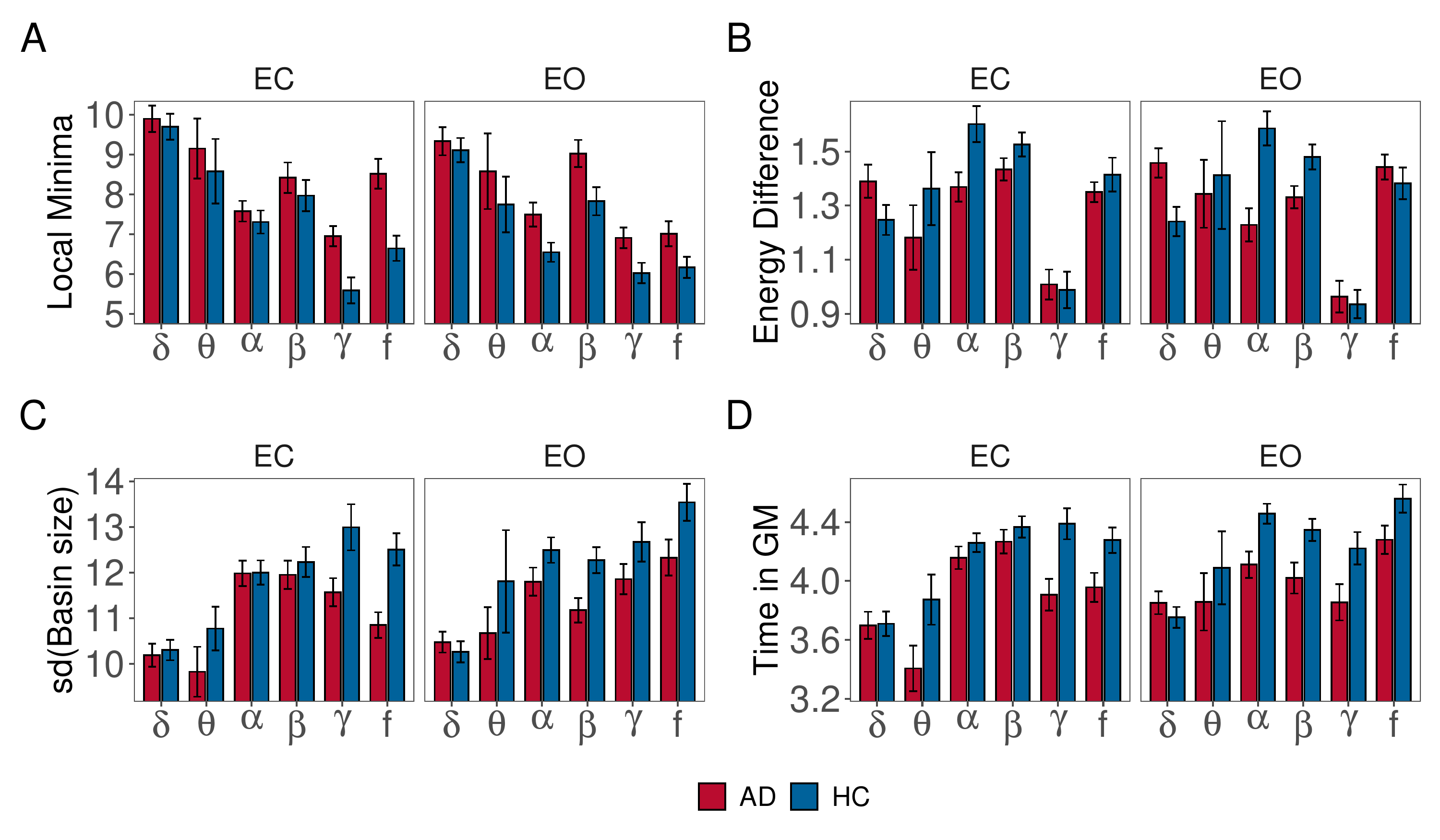}
    \caption{Features extracted from EL. Bars show mean value and errorbars show the 95\% confidence interval of mean of:
    A) Number of LM
    B) Mean energy difference between LM and GM
    C) Standard deviation of basin sizes
   D) Simulated duration in the basin of GM}
    \label{fig:results}
\end{figure*}

The number of LMs was computed for each participant, band and condition (Fig. \ref{fig:results}A). AD patients have consistently more LMs than HC $(F(1,1440)=43.32,p<0.001)$. Moreover, significant main effects of band $(F(5,1440)=56.47,p<0.001)$ and condition $(F(1,1440)=6.55,p=0.011)$ were found. None of the interactions are significant: group:band $(F(5,1440)=1.64,p=0.146)$, group:condition $(F(1,1440)=0.077,p=0.781)$, band:condition $(F(5,1440)=2.189,p=0.053)$ and group:band:condition $(F(5,1440)=1.211,p=0.302)$.

The average energy difference between the GM and the rest of the LMs was computed for each participant, band and condition (Fig. \ref{fig:results} B). There is a significant main effect of band $(F(5,1440)=41.265,p<0.001)$. No significant between-group $(F(1,1440)=2.049,p=0.153)$ or between-condition $(F(1,1440)=0.919,p=0.338)$ effects were found. There is, however, a significant interaction between group and band $(F(5,1440)=8.286,p<0.001)$. The rest of the interactions are not significant: group:condition $(F(1,1440)=0.559,p=0.455)$, band:condition $(F(5,1440)=0.827,p=0.531)$, group:band:condition $(F(5,1440)=1.04,p=0.392)$. The significant group:band interaction was analyzed. AD have significantly higher average energy difference in $\delta$ band $(F(1,277)=10.153,p=0.001)$. HC have significantly higher average energy difference in $\alpha$ $(F(1,277)=22.705,p<0.001)$ and $\beta$ $(	
F(1,277)=7.684,p=0.024)$.

For each participant, condition and band, the sizes of all basins of LMs are computed and their standard deviation is calculated. These values are then square-root transformed in order to be normally distributed.

AD show consistently smaller standard deviation of basin size than HC $(F(1,1440)=39.339,p<0.001)$ (Fig. \ref{fig:results} C). Moreover, significant main effects of band $(F(5,1440)=29.552,p<0.001)$ and condition $(F(1,1440)=4.548,p=0.033)$ were identified. The group:band $(F(5,1440)=3.344,p=0.005)$ and band:condition $(	F(5,1440)=3.155,p=0.008)$ interactions were significant. On the other hand, the group:condition $(F(1,1440)=0.031,p=0.86)$ and group:band:condition $(F(5,1440)=1.005,p=0.413)$ interactions do not show significant differences.

The significant group:band interaction was tested. Significant differences were found in $\theta$ $(F(1,277)=25.65,p<0.001)$, $\beta$ 	
$(F(1,277)=5.273,p=0.022)$, $\gamma$ $(F(1,277)=7.705,p=0.006)$ bands and in full frequency $(F(1,277)=15.215,p<0.001)$ wherein AD have lower standard deviation of basin size than HC. No significant differences were found in $\delta$ $(F(1,277)=0.042,p=0.837)$ and $\alpha$ $(F(1,277)=1.577,p=0.21)$.

Testing of the band:condition interaction shows that the standard deviation of basin size is significantly higher during EO condition in full frequency $(F(1,277)=11.419,p=0.001)$. In the remaining bands there are no significant differences: $\delta$ $(F(1,277)=0.259,p=0.611)$, $\theta$ $(F(1,277)=3.622,p=0.058)$, $\alpha$ $(F(1,277)=0.332,p=0.565)$, $\beta$ $(F(1,277)=1.393,p=0.239)$ and $\gamma$ $(F(1,277)=0.002,p=0.964)$.

20000 samples of state-transitions were simulated for each participant, condition and band and the time spent within the basin of GM was calculated.

AD remain within the GM basin significantly shorter time than HC $(F(1,1440)=46.898,p<0.001)$ (Fig. \ref{fig:results}D). Significant main effect of band was also found $(F(5,1440)=26.878,p<0.001)$. There are also significant interactions of group:band $(F(5,1440)=3.433,p=0.004)$ and band:condition $(F(5,1440)=3.458,p=0.004)$. No significant effect of condition $(F(1,1440)=3.143,p=0.076)$ was found. The remaining interactions were not significant: group:condition 1.47e-05	$(F(1,1440)=0.024,p=0.876)$ and group:band:condition $(F(5,1440)=1.016,p=0.407)$.

Tests of group:band show shorter time in GM basin for AD in $\theta$ 	
$(F(1,277)=21.295,p<0.001)$, $\alpha$ $(F(1,277)=7.984,p=0.005)$, $\beta$ $(F(1,277)=5.56,p=0.019)$, $\gamma$ $(F(1,277)=13.772,p<0.001)$ and full frequency $(F(1,277)=10.31,p=0.001)$. No significant difference was found in $\delta$ $(F(1,277)=0.449,p=0.503)$.

Tests of band:condition interaction reveal longer duration in GM basin during EO condition in $\theta$ $(F(1,277)=4.026,p=0.046)$ and full frequency $(F(1,277)=11.41,p=0.001)$. No significant difference was found in $\delta$ $(F(1,277)=0.927,p=0.337)$, $\alpha$ $(F(1,277)=1.524,p=0.218)$, $\beta$ $(F(1,277)=1.933,p=0.166)$ and $\gamma$ $(F(1,277)=0.655,p=0.419)$.

Finally, we observed  significant correlations between energy of brain states and MMSE (Fig \ref{fig:correlation}). For each combination of band and condition we looked only at the strongest significant correlation. Specifically, significant correlation was found in EC condition in $\theta$ ($\tau=0.6, p=0.01$) and $\alpha$ ($\tau=-0.51, p=0.01$) and full ($\tau=0.49, p=0.01$), and in EO condition in $\delta$ ($\tau=-0.51, p=0.01$), $\theta$ ($\tau=0.59, p=0.01$), $\alpha$ ($\tau=-0.43, p=0.03$) and full ($\tau=-0.49, p=0.01$).

\begin{figure*}[!ht]
    \centering
    \includegraphics[width=0.8\linewidth]{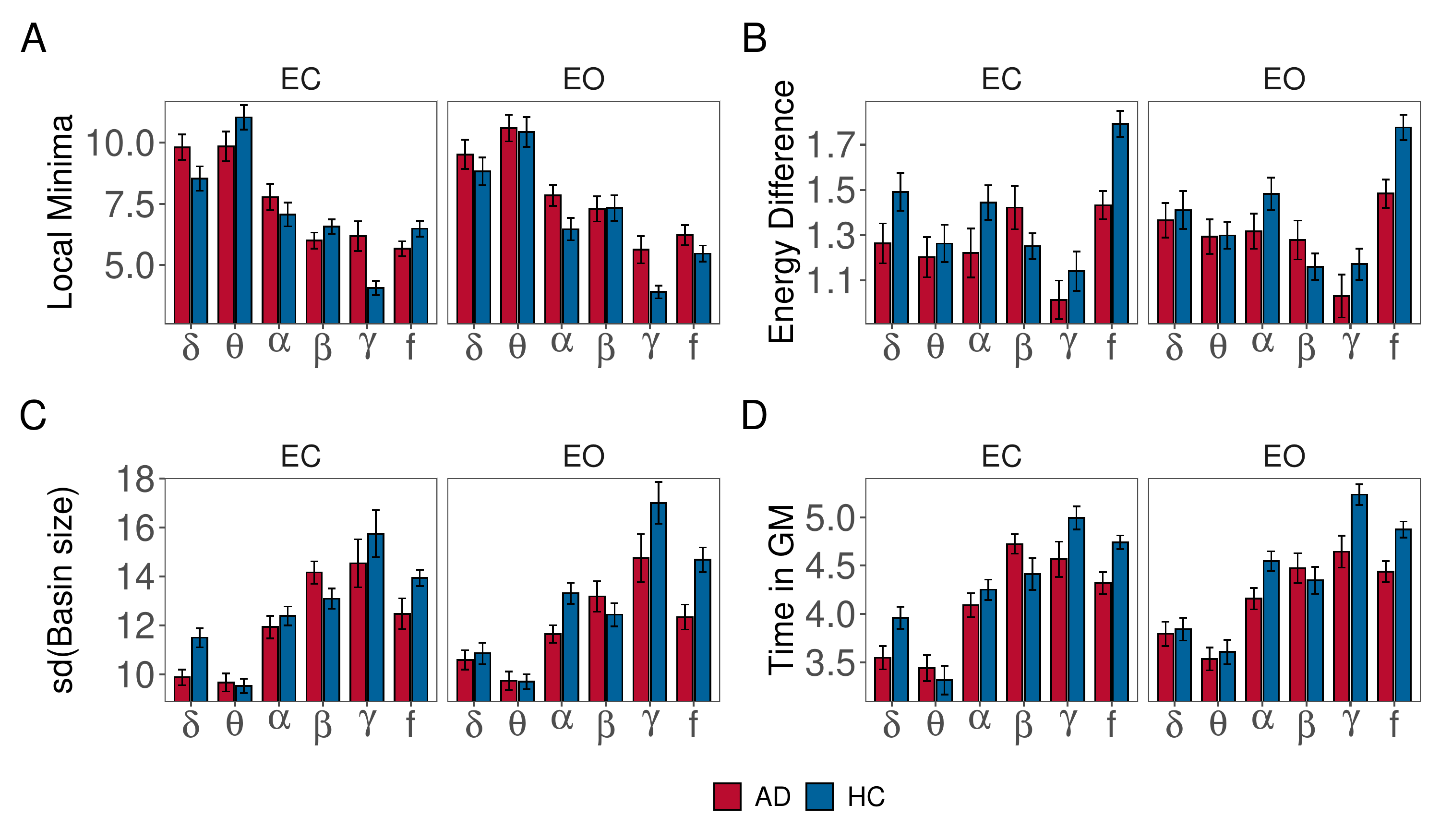}
    \caption{Features extracted from EL using the data from participants above 70 years.}
    \label{fig:results-above70}
\end{figure*}

\begin{figure}[tb]
    \centering
    \includegraphics[width=\linewidth]{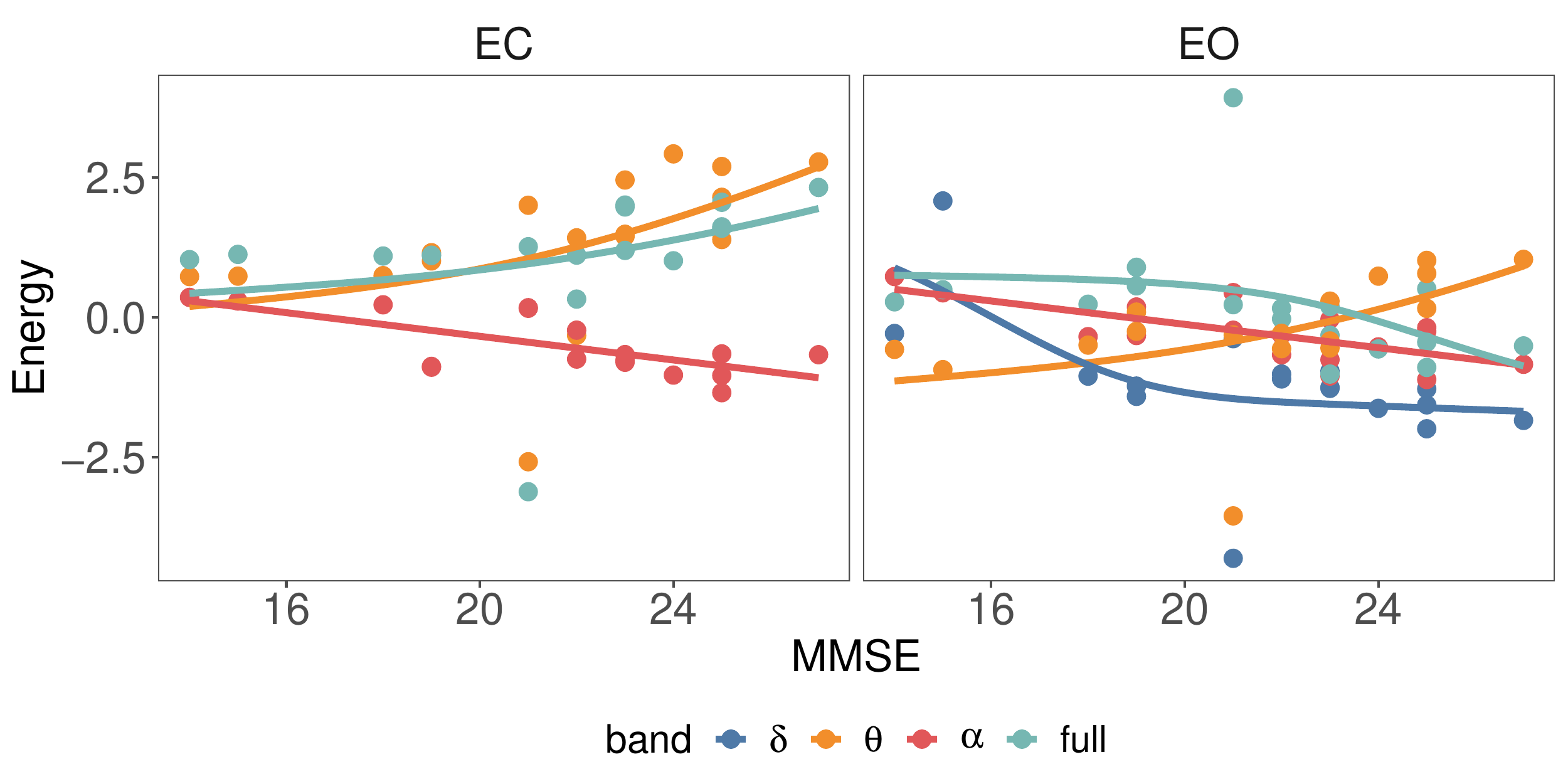}
    \caption{Relationships between energy of states and cognitive scores (MMSE).}
    \label{fig:correlation}
\end{figure}

\section{Discussion and conclusions}
In this study, for the first time, an EL approach is applied to study changes in global dynamics of EEG from patients with AD. Furthermore, we have extended the traditional process of estimating the EL to an entirely data-driven approach and proposed adjustments to apply the method to EEG.

The data-driven channel selection approach allows estimating ELs by selecting a subset of channels without relying on any prior assumptions about the regions of interest. The channels selected by our approach (Fig. \ref{fig:selection}) are mainly in the central, parietal and temporal areas of the scalp. The frontal regions  were rarely selected. The selected areas are in line with the findings of slowing of EEG rhythms in AD, such as the increase of $\theta$ power and decrease of $\beta$ power in parieto-occipital regions \cite{jeong2004slowing}. 

We show that using EL is worth the additional computation cost as models trained on energy values results in significantly better performance compared to models trained on pairwise functional connectivity (Fig. \ref{fig:energy-connectivity}). Moreover, this suggests that the EEG signals cannot be fully explained as a product of a weighted network. 

Interestingly, several significant differences between conditions and frequency bands were found both in terms of the performance of the machine learning models and the features of EL. Since the presented results are the first application of this method to EEG data, we cannot interpret the results in the context of previous research, and thus we focus mainly on the differences between groups. The differences between conditions are rather predictable as the brain networks are presumably in two distinct configurations during EC and EO conditions. Similarly, the differences between bands suggest that each band encodes different information, which is supported by the findings of different roles of frequency components of EEG \cite{cohen2017bandroles}.

Comparison of the ELs revealed that AD cases have consistently more LMs than HC (Fig. \ref{fig:results}A). The LM can be viewed as attractors in the state space since it is a state with a high probability of occurrence. To interpret the meaning of such finding in the context of previous research, the LM should be likened to the constraints of the system. It is well described that a balanced and temporally precise pattern of synchronisation and desynchronisation is pertinent to cognitive function \cite{schnitzler2005}. The constraints of the system revealed in this work might reflect the inability of AD networks to show such degree of flexibility. The EL is, therefore quantification of constraints of a multivariate system. Thus, increased complexity of EL means a decrease in complexity of the underlying signals, which fits well with the findings of previous research showing a decrease in complexity in single channel and pairs of channels \cite{ahmadlou2010complexity,garn2015complexity,coronel2017complexity,jeong2004slowing}. In other words, the brains of AD patients have fewer degrees of freedom. This finding might be correlated with the early pathology of AD, i.e. the loss of synaptic contacts, correlated to various cognitive decrements \cite{terry2000celldeath}. We  speculate that the loss of synapses and/or neurons and the subsequent decline in communication between cortical regions \cite{jeong2004slowing} might lead to the observation of more LMs, i.e. constraints, in addition to the "normal" LMs which probably have a functional role. We speculate that these extra LM in AD do not have any function; instead, they might be responsible for the disruption of information processing \cite{jeong2004slowing, jeong2001mutual}.

The GM largely determines the shape of the EL as it is an attractor with a large field of attraction. The difference in energy between the GM and the remaining LMs then quantifies the ease of transition between them. We calculated the average value of these differences (Fig. \ref{fig:results}B). In $\delta$ band AD have higher energy difference which indicates easier state transitions. Research shows that the power in $\delta$ band increases in AD \cite{jeong2004slowing,brenner1988slowing,ghorbanian2015slowing}; similarly, our finding allows less organised state transitions, i.e. more activity. Average energy difference of AD is lower in $\alpha$ and $\beta$, which could be linked to the decrease in these bands \cite{jeong2004slowing,brenner1988slowing,ghorbanian2015slowing}.

The GM largely defines the variation in the sizes of basins of LMs. This is because the GM basin is always the largest, which means that calculating the variation of basin sizes essentially measures the similarity of LMs to the GM in terms of basin size. Our analysis reveals that AD cases have a smaller variation of basin sizes, indicating higher similarity to the GM (Fig. \ref{fig:results}C). This result needs to be interpreted together with the number of LM. This is given by the definition of how basin size is computed, as each state belongs to one and only one basin. This, in turn, means that if an EL has more LM, then their basins tend to be smaller since the number of states remains constant.  It can then be concluded that AD cases have smaller basins which are similar to each other. Smaller basins also mean that the states in them are more strongly attracted to their LMs since they are closer to the centre of the basin. We interpret this as LMs of AD being stronger constraints compared to HC thus supporting the account of AD, leading to decreased complexity of signals \cite{coronel2017complexity,garn2015complexity,ahmadlou2010complexity,jeong2004slowing}.

Finally, the simulated duration in the basin of a GM was analysed. We focus only on the GM, whereas with size, the attractor strength declines - states on the edges of the basin are less affected by the GM than states close to the centre. We demonstrate that AD cases spend a significantly shorter time in the basin of GM (Fig. \ref{fig:results}D). Since AD cases have more LMs that are similar to the GM, the time spent within each basin must be divided evenly among all LMs. In other words, a large proportion of states that the AD cases must "visit" is predetermined by the LMs. On the other hand, HC cases are constrained mainly by a large but weak basin of the GM and a few additional strong LMs (as evidenced by the higher variation in basin sizes). This result again reinforces the finding that the signals of AD cases are more constrained, i.e. less complex than HC cases.

By analysing data from participants above 70, we demonstrate that the reported differences in EL are transferable across data. Due to the age difference between the datasets, we kept these analyses separate. Fig 6 shows the differences observed in the above 70 dataset are generally comparable to those of under 70 but seem to be smaller or disappear entirely in few cases. We speculate that this is caused by age differences.

Finally, we showed that EL correlates strongly with the MMSE cognitive score of AD patients (Fig 7) as all reported absolute correlations are stronger than 0.5, with the strongest one occurring in the full band during EO condition. This suggests that EL has the potential to be used as a biomarker of the degree of cognitive deficits in AD, i.e. predicting symptom severity in addition to being a diagnostic aid, thus showing potential in allowing disease progression monitoring and response to treatment in large pharmaceutical trials. In the current study, we did not analyse which brain states correlate with MMSE, which should be addressed in future work. However, these states are not LMs, thus suggesting that future research should aim to design methods for analysing the whole EL instead of prioritising the importance of LMs.

The reported research has some limitations. First, we achieved  only medium accuracy of the pMEM estimates. A possible cause is the length of the EEG recordings, which might be too short to obtain higher accuracy. Second, the data need to be binarised by using a threshold \cite{ezaki2018landscape}. Such simplification leads to considerable loss of information as compared to the continuous EEG signals. We attempt to lessen the impact of binarisation by using multiple thresholds instead of one. While using multiple thresholds does not resolve the stationarity issue, it seems to capture a larger portion of information since it improves the predictive accuracy (Fig. \ref{fig:selecting-parameters}). 

Third, fitting pMEM is computationally expensive, and currently, it is only possible to estimate the EL with 10-15 channels \cite{ezaki2018landscape} as the number of possible states increases exponentially with more channels. Previous research relied on manually selecting regions of interest which might lead to biased selections. Our data-driven channel selection resolves the issue. On the other hand, such channel selection produces different selections for each band, making band-wise comparisons difficult.

\section*{Acknowledgement}
The original data collection was funded by a grant from the Alzheimer’s Research UK (ARUK-PPG20114B-25). The views expressed are those of the author(s) and not necessarily those of the NHS, the NIHR or the Department of Health.

%\newpage
\bibliographystyle{unsrt}
\bibliography{ref}

\end{document}